\documentclass[aps,superscriptaddress]{revtex4}
\usepackage{amssymb,amsmath,epsfig}
\usepackage[colorlinks=true, pdfstartview=FitV, linkcolor=blue, citecolor=red, urlcolor=magenta, breaklinks=true]{hyperref}
\usepackage{subfigure}
\usepackage{amsfonts}
\usepackage{graphicx,epsfig} 
\usepackage{subfigure}

\begin{document}
\title{The generalized uncertainty principle effect in acoustic black holes}

\author{M. A. Anacleto}\email{anacleto@df.ufcg.edu.br}
\affiliation{Departamento de F\'{\i}sica, Universidade Federal de Campina Grande
Caixa Postal 10071, 58429-900 Campina Grande, Para\'{\i}ba, Brazil}

\author{F. A. Brito}\email{fabrito@df.ufcg.edu.br}
\affiliation{Departamento de F\'{\i}sica, Universidade Federal de Campina Grande
Caixa Postal 10071, 58429-900 Campina Grande, Para\'{\i}ba, Brazil}
\affiliation{Departamento de F\'isica, Universidade Federal da Para\'iba, 
Caixa Postal 5008, 58051-970 Jo\~ao Pessoa, Para\'iba, Brazil}

\author{G. C. Luna} \email{gabrielacluna@hotmail.com}
\affiliation{Departamento de F\'{\i}sica, Universidade Federal de Campina Grande
Caixa Postal 10071, 58429-900 Campina Grande, Para\'{\i}ba, Brazil}

\author{E. Passos}\email{passos@df.ufcg.edu.br}
\affiliation{Departamento de F\'{\i}sica, Universidade Federal de Campina Grande
Caixa Postal 10071, 58429-900 Campina Grande, Para\'{\i}ba, Brazil}

\begin{abstract} 
We obtain an effective acoustic metric with quantum corrections that are provided by a minimum length implemented by the generalized Heisenberg uncertainty principle (GUP) in the Abelian Higgs model. 
The effective acoustic metric now depends on the contribution of scalar and vector potentials.
We also explore the Hawking radiation and entropy by considering the effective canonical acoustic black hole and find that the modified Hawking temperature leads to logarithm corrections to the entropy. Finally, we investigate the dispersion relations of the model to establish the relationships among the deviations of the group velocity, frequency and temperature due to the GUP.
\end{abstract}

\maketitle
\pretolerance10000

\section{Introduction}
Black holes are fascinating objects in our universe and an astrophysical observation of these objects is a subject of great interest in General Relativity.
Recently, the LIGO-VIRGO collaboration~\cite{LIGOScientific:2016aoc,LIGOScientific:2017vwq} with the remarkable observation of GW150914 made its first direct detection of gravitational waves 
that emerge from the merger of a binary black hole system.
In addition, the Event Horizon Telescope captured the image of a supermassive black hole at the center of galaxy M87~\cite{event2019firstI,event2019firstVI}.
Thus, with these two observations that evidence the existence of black holes, a wide range appears to deepen studies in the development of black hole physics.

In recent decades great advances have been made in the physics of black holes in the investigation of properties near the event horizon.
One of the characteristics of black holes is Hawking radiation.
From an observational point of view, the detection of Hawking radiation is difficult due to technical limitations.
To overcome this difficulty, Unruh~\cite{Unruh:1980cg,Unruh:1994je}  in 1981 introduced the acoustic black hole model that makes possible to detect Hawking radiation from the acoustic black hole in the laboratory. Thus, several works have been done to analyze analogous Hawking radiation in different physical scenarios~\cite{Visser:1997ux}(see \cite{Barcelo:2005fc} for review).
In~\cite{Vieira:2014rva} the analytical solutions of the massless scalar field and the analogous Hawking radiation were addressed.

Furthermore, starting with the Abelian Higgs model, relativistic metrics of acoustic black holes have been constructed~\cite{Ge:2010wx,Anacleto:2010cr,Anacleto:2011bv,Anacleto:2013esa} 
(see also \cite{Bilic:1999sq,Fagnocchi:2010sn,Visser:2010xv}). 
In~\cite{Ge:2019our}, by considering relativistic Gross-Pitaevskii theory and Yang-Mills theory, the acoustic black hole in curved geometry has been obtained. 
In addition, based on a holographic approach, the acoustic black hole constructed from  black D3-brane has been discussed in~\cite{Yu:2017bnu}.
Moreover, by applying these effective metrics, studies have been carried out to examine the phenomena of superradiance~\cite{Basak:2002aw,Richartz:2009mi,Anacleto:2011tr,Zhang:2011zzh,Ge:2010eu}, entropy~\cite{Zhao:2012zz,Anacleto:2014apa,Anacleto:2015awa,Anacleto:2016qll,Anacleto:2019rfn}, quasinormal modes~\cite{Cardoso:2004fi,Nakano:2004ha,Berti:2004ju,Chen:2006zy,Guo:2020blq,Ling:2021vgk} and the analogous Aharonov-Bohm effect~\cite{Dolan:2011zza,Anacleto:2012ba,Anacleto:2012du,Anacleto:2015mta,Anacleto:2016ukc,Anacleto:2018acl} 
as well as quantum stochastic motion~\cite{Anacleto:2020kxj,Anacleto:2021wmv}. 
In~\cite{Qiao:2021trw} the gravitational bending of acoustic Schwarzschild black hole was investigated.
Recently, great progress has been made in detecting Hawking radiation 
in the laboratory~\cite{MunozdeNova:2018fxv,Isoard:2019buh}.
A first experimental realization of the acoustic black hole in the Bose-Einstein condensate (BEC) has been reported in~\cite{Lahav:2009wx}.
In addition, experimental studies have been carried out 
on optical systems~\cite{Steinhauer:2014dra,Drori:2018ivu,Rosenberg:2020jde}
and also on other physical systems~\cite{Guo:2019tmr,Bera:2020doh,Blencowe:2020ygo}.
 
The aim of this paper is to find a metric with quantum corrections for the acoustic black hole. 
For this purpose, we start with the Abelian Higgs model and apply the same procedure done in~\cite{Ge:2010wx,Anacleto:2010cr,Anacleto:2011bv,Anacleto:2013esa}.
It has been obtained in~\cite{Quesne:2006fs} generalized commutation relation that arises due to the presence of a minimum length given by
\begin{eqnarray}
[X^{\mu},P^{\nu}]=-i[(1-\beta p^{\sigma}p_{\sigma})g^{\mu\nu}-2\beta p^{\mu}p_{\nu}],
\end{eqnarray}
where $ \beta $ is the deformation parameter with minimum length of the order of $ \sqrt{\beta} $. 
This naturally leads to the generalized Heisenberg uncertainty principle (GUP)~\cite{Garay:1994en,AmelinoCamelia:2000ge,Ali:2009zq,Das:2008kaa,Das:2009hs,Ali:2011fa,Scardigli:1999jh,Capozziello:1999wx,Scardigli:2003kr,Tawfik:2014zca,Dahab:2015,Dutta:2015, KMM,Gangopadhyay:2013ofa}. As shown in~\cite{Quesne:2006fs,Dias:2016lkg,HoffDaSilva:2020uov} the above commutation relation holds  in the configuration space for $ P^{\mu}=(1-\beta p^{\sigma}p_{\sigma})p^{\mu} $, and then a correction for the partial derivative operator is obtained, i.e., 
$ \partial_{\mu}\rightarrow (1+\beta\Box)\partial_{\mu} $.
Thus, in order to generate the effective acoustic metric, we incorporate the contribution of the minimum length with this prescription into the Lagrangian.
In recent years, the exploration of models considering the GUP effect has been a subject of great interest in the literature~\cite{Yang:2018wlb,Feng:2017gms,Feng:2015jlj,Javed:2019btd,Li:2016mwq,Chen:2016ftz,Maluf:2018lyu,Gomes:2018oyd,Pourhassan:2018chj,Sadeghi:2016xym,Scardigli:2014qka,Casadio:2020rsj,Scardigli:2016pjs,Scardigli:2019pme,Anacleto:2021qoe}.
In particular, the analysis of the effect of the GUP on Hawking radiation has been an issue that has been widely explored by many authors~\cite{Sakalli:2016mnk,Gecim:2017zid,Alonso-Serrano:2018ycq,Anacleto:2015mma,Silva:2012mt,Anacleto:2015rlz,Anacleto:2015kca,Haldar:2018zyv,Haldar:2019fcz,Iorio:2019wtn,Scardigli:2008jn,Anacleto:2020efy,Anacleto:2020zfh,Nozari:2005ah,Nozari:2008gp,Nozari:2008rc,Nozari:2006ka}. Therefore, the obtained metric presents quantum corrections that are identified by the deformation parameter. As a result, we show that terms associated with scalar and vector potentials appear in the effective metric of the acoustic black hole due to the deformation parameter $ \beta $.

The paper is organized as follows. In Sec.~\ref{GUP} we derive the acoustic metric under the GUP. In Sec.~\ref{dispersion} we investigate the effect of the GUP in the dispersion relations to establish relationships among deviations on the group velocity, frequency and Hawking temperature. In the Sec.~\ref{conclu} we make our final comments.

\section{GUP-Corrected Acoustic Metric}
\label{GUP}

{In this section, we consider the Abelian Higgs model that describes high energy physics in order to derive a GUP-corrected acoustic black hole metric. 
This study is motivated by the fact that acoustic black holes could be formed in high energy physical processes, such as quark matters, neutron stars and quark gluon plasma. 
Thus, it seems to be natural to look for acoustic black holes in a quark gluon plasma fluid with GUP correction in this regime. 
In addition acoustic phenomena in quark gluon plasma 
matter was investigated in~\cite{Casalderrey-Solana:2004fdk} and acoustic black holes 
in a plasma fluid was explored in~\cite{GarciadeAndrade:2008kz}.}
Thus, we start with the following Lagrangian of the Abelian Higgs model: 
\begin{eqnarray}
\mathcal{L}=-\frac{1}{4}F_{\mu\nu}F^{\mu\nu}+\left\vert\left(\partial_{\mu}-ieA_{\mu}\right)\phi\right\vert^{2}+m^{2}\vert\phi\vert^{2}-b\vert\phi\vert^{4},
\label{higgs}
\end{eqnarray}
{where $ A_\mu $ is the gauge field 
and $F_{\mu\nu}=\partial_{\mu}A_{\nu}-\partial_{\nu}A_{\mu}$ is the field strength tensor. The model deals with a charged fluid since the Lagrangian is invariant under $U(1)$ gauge symmetry
which allows a minimal coupling between the complex scalar field $ \phi $ and the gauge field $A_\mu$  via Noether current and coupling constant $e$.}

Let us first consider only the free scalar field sector given by 
\begin{eqnarray}
\label{ef}
\mathcal{L}=\left\vert\partial_{\mu}\phi\right\vert^2+m^{2}\vert\phi\vert^{2}.
\end{eqnarray}
In order to obtain an effective acoustic metric, 
we introduce the contribution of the minimum length in the above Lagrangian by means of the following prescription~\cite{Quesne:2006fs,Dias:2016lkg,HoffDaSilva:2020uov}:
\begin{eqnarray}
\partial_\mu \rightarrow \left( 1+\beta \square \right) \partial_\mu ,
\label{dpart}
\end{eqnarray}
where $ \beta $ is the deformation parameter.
By substituting (\ref{dpart}) into (\ref{ef}), we have
\begin{eqnarray}
\label{lmodf}
\mathcal{L}=\left\vert\partial_{\mu}\phi\right\vert^2+2\beta\square\left\vert\partial_{\mu}\phi\right\vert^2 +m^2\vert\phi\vert^2.
\end{eqnarray}
Here we keep the terms up to first order in $\beta$. 
Thus, we obtain the modified Klein-Gordon equation given by
\begin{eqnarray}
\square\phi +2\beta\square ^2 \phi -m^2\phi=0 .
\label{eq2}
\end{eqnarray}
Now isolating the term $\square\phi$ in (\ref{eq2}) and replacing it in (\ref{dpart}), we find
\begin{eqnarray}
\mathcal{L}=\left\vert\left( 1+\beta m^2\right) \partial_\mu\phi\right\vert^2+m^2\vert\phi\vert^2 .
\end{eqnarray}
So the new contribution of the derivative is 
$\partial_\mu\rightarrow\left( 1+\beta m^2\right)  \partial_\mu$. 
Now we replace this new contribution of the derivative in the Lagrangian (\ref{higgs}) and so we get
\begin{eqnarray}
\mathcal{L}=-\frac{1}{4}F_{\mu\nu}F^{\mu\nu}(1+2\beta m^2)+\left\vert\left[ \left(1+\beta m^2\right) \partial_{\mu}-ieA_{\mu}\right]\phi\right\vert^{2}+m^{2}\vert\phi\vert^{2}-b\vert\phi\vert^{4}.
\end{eqnarray}
Using the Madelung representation for the scalar field
$\phi = \sqrt{\rho\left( \vec{x},t\right) }e^{iS\left( \vec{x},t\right) } $, 
the Lagrangian becomes
\begin{eqnarray}
\mathcal{L}&=&-\frac{1}{4}F_{\mu\nu}F^{\mu\nu}(1+2\beta m^2)
+\frac{1}{4\rho}\partial _\mu \rho \partial ^\mu \rho \left( 1+2\beta m^2\right)
+\left( 1+2\beta m^2\right)\rho\partial _\mu S\partial^\mu S 
\nonumber\\
&-&2\left( 1+\beta m^2\right) e\rho A^{\mu} \partial_\mu S+e^2 A_\mu A^\mu \rho +m^2 \rho -b\rho ^2,
\end{eqnarray}
and the equations of motion for $ S $ and $ \rho $ are
\begin{eqnarray}
 \partial_\mu\left[\rho(\partial ^\mu S-eA^\mu)\left( 1+2\beta m^2\right)
+\beta m^2 e\rho A^{\mu}\right] =0,
\label{mov1}
\end{eqnarray}
and
\begin{eqnarray}
\frac{1}{\sqrt{\rho}}\partial _\mu \partial ^\mu\sqrt{\rho}\, (1+2\beta m^2) 
-\left( \partial _\mu S-e A_\mu\right)^2 (1+2\beta m^2)
-2\beta m^2 \left( \partial _\mu S-e A_\mu\right) e A^\mu 
- m^2+2b\rho =0.
\label{mov2}
\end{eqnarray}
{The Eq.~(\ref{mov1}) is the continuity equation and Eq.~(\ref{mov2}) is an equation describing a hydrodynamical fluid 
with a quantum potential term $ \frac{1}{\sqrt{\rho}}\partial _\mu \partial ^\mu\sqrt{\rho}\, (1+2\beta m^2) $ which can be negligible in the hydrodynamic region.
However, in this regime the model still describes relativistic fluids.
On the other hand, one can also apply the Madelung representation of the condensate wave function in Abelian Higgs model and in relativistic Bose-Einstein condensate as has been argued in \cite{Ge:2010wx} and \cite{Fagnocchi:2010sn,Giacomelli:2017eze}, respectively. 
Thus, just as in the Schroedinger equation, this approach yields the continuity and the Euler equation applied to an irrotational and inviscid fluid in the absence of the quantum potential.}

For the gauge field $ A_{\mu} $, we find
\begin{eqnarray}
\partial_{\mu}F^{\mu\nu}(1+2\beta m^2)=2e\rho(1+\beta m^2)u^{\nu} + 2\beta m^2 e^2\rho A^{\nu},
\end{eqnarray}
or
\begin{eqnarray}
\partial_{\mu}F^{\mu\nu}=2e\rho(1-\beta m^2)u^{\nu} + 2\beta m^2 e^2\rho A^{\nu},
\end{eqnarray}
where $ u^\mu=\partial ^\mu S-eA^\mu=(-\omega,-v^i) $.

Next we make the following perturbation in the equations of motion \eqref{mov1} and \eqref{mov2}:
\begin{eqnarray}
&&S=S_0+\epsilon S_1 + \mathcal{O}(\epsilon ^2), \\
&&\rho = \rho _0+\epsilon \rho _1 + \mathcal{O}(\epsilon ^2),
\end{eqnarray}
so that we obtain
\begin{eqnarray}
\partial _\mu \left\lbrace \rho _1 (1+2\alpha) u_0^{\mu}
+\rho _1\alpha e A^\mu 
+\rho _0\left( 1+2\alpha\right) \partial ^\mu S_1\right\rbrace  =0,
\label{13}
\end{eqnarray}
and
\begin{eqnarray}
(1+2\alpha) u_0^{\mu} \partial _\mu S_1
+ \alpha e A^{\mu}\partial_\mu S_1 -b\rho _1 =0.
\label{15}
\end{eqnarray}
For simplicity we have defined $ u_0^\mu =\partial ^\mu S_0-eA^\mu $ and 
 $\alpha=\beta m^2$ {is the dimensionless GUP parameter.}
Now, solving \eqref{15} for $\rho _1$ and substituting into \eqref{13}, we find
\begin{eqnarray}
\partial _\mu 
\left[u^{\mu}_0u^{\nu}_0(1+4\alpha) +\alpha eu^{\mu}_0 A^{\nu}+\alpha e A^{\mu}u^{\nu}_0
+ b\rho _0 \left( 1+2\alpha\right)g^{\mu\nu}\right]\partial_\nu S_1  =0.
\end{eqnarray}

We can also write the above equation as follows:
\begin{eqnarray}
\label{eqwave}
&&\partial _t\left\lbrace  \tilde{\omega}^2_0 
\left[ -1-4\alpha +\frac{2\alpha e A_t}{\tilde{\omega} _0}- \frac{b\rho_0}{\tilde{\omega}^2_0}\left( 1+2\alpha\right) \right]\partial _t S_1 
+\tilde{\omega}^2_0\left[ -\frac{v^i_0}{\tilde{\omega}_0}\left(1+4\alpha - \frac{e A_t}{\tilde{\omega}_0}\right)
+\alpha\frac{eA^i}{\tilde{\omega} _0} \right]\partial _i S_1 \right\rbrace 
\nonumber\\
&& +\partial _i\left\lbrace \tilde{\omega}^2_0 
\left[-\frac{v^i_0}{\tilde{\omega}_0}\left(1+4\alpha - \frac{e A_t}{\tilde{\omega}_0}\right)
+\alpha\frac{eA^i}{\tilde{\omega} _0}  \right]\partial _t S_1 \right.
\nonumber\\
&&+\left.\tilde{\omega}^2_0\left[ - (1+4\alpha)\frac{ v^i_0 v^j_0}{\tilde{\omega}^2_0}+\alpha \left( v^i\frac{e A^j}{\tilde{\omega} _0}+v^j\frac{eA^i}{\tilde{\omega} _0}\right) +\frac{b\rho_0}{\tilde{\omega}^2_0}\left( 1+2\alpha\right) \delta^{ij}\right]\partial _j S_1 \right\rbrace  =0,
\end{eqnarray}
where $\tilde{\omega} _0=eA_t+\omega_0$, $ \omega_0=-\partial^t S_0 $ and $v_0^i=\partial _i S_0+eA^i$ (the local velocity field).
In addition, at the weak field limit, we define 
$ c^2_s=b\rho_0/\omega^2_0$,  
{$ v^i=v^i_0/\omega_0 $, 
$ \Phi=A_t/\omega_0 $ and $ \Lambda^i=A^i/\omega_0 $.  In order to identify an acoustic metric, the last three quantities should be recognized as the vector velocity, electric scalar potential and magnetic vector potential, respectively.}
So the equation \eqref{eqwave} becomes
\begin{eqnarray}
\label{KG}
&&\partial _t\left\lbrace \frac{b\rho _0}{c_s ^2}
\left[ -1-4\alpha +2\alpha e\Phi -c_s^2\left( 1+2\alpha\right) \right] \partial _t S_1 
+ \frac{b\rho _0}{c_s ^2}\left[ -(1+4\alpha-\alpha e\Phi) v^i +\alpha e\Lambda^i\right] \partial _iS_1 \right\rbrace 
\nonumber\\
&& +\, \partial _i\left\lbrace \frac{b\rho _0}{c_s^2}
\left[ -(1+4\alpha-\alpha e\Phi) v^i +\alpha e\Lambda^i\right]\partial _t S_1 \right.
\nonumber\\
&&+\left.\frac{b\rho _0}{c_s ^2}\left[ -(1+4\alpha) v^i v^j+\alpha e\Lambda ^j v^i 
+\alpha v^j e\Lambda ^i +c_s^2\left( 1+2\alpha\right) \delta^{ij}\right]
\partial _j S_1 \right\rbrace =0.
\end{eqnarray}
We can now write the above equation as the equation of the fluctuations in a (3+1) dimensional curved space as follows:
\begin{eqnarray}
\frac{1}{\sqrt{-g}}\partial_{\mu}\left( \sqrt{-g}g^{\mu\nu}\partial_{\nu}\right)\psi=0,
\end{eqnarray}
where
\begin{eqnarray}
\sqrt{-g}g^{\mu\nu}=\frac{b\rho_0}{c^2_s}
\left(\begin{array}{ccc}
-1-c^2_s(1+2\alpha) - 2\alpha (2- e\Phi) & \vdots & -v^{i}(1+4\alpha -\alpha e\Phi)+\alpha e\Lambda^i \\ 
\cdots\cdots &\cdot &\cdots\cdots\\
-v^{j}(1+4\alpha -\alpha e\Phi)+\alpha e\Lambda^j & \vdots & c^2_s(1+2\alpha)\delta^{ij} - v^i v^j (1+4\alpha)+ \alpha v^i e\Lambda^j 
+\alpha e\Lambda^i v^j
\end{array} 
\right).
\end{eqnarray}
Therefore, the effective relativistic acoustic metric is given by
\begin{eqnarray}
\label{metr}
g_{\mu\nu}=\frac{b\rho_0}{c_s\sqrt{\mathcal{Q}}} 
\left(\begin{array}{ccc}
-c^2_s(1+2\alpha) + v^2(1+4\alpha)- 2\alpha e\vec{v}\cdot\vec{\Lambda} & \vdots & -v^{i}(1+4\alpha 
-\alpha e\Phi)+\alpha e\Lambda^i \\ 
\cdots\cdots &\cdot &\cdots\cdots\\
-v^{j}(1+4\alpha -\alpha e\Phi)+\alpha e\Lambda^j & \vdots & [1+c^2_s(1+2\alpha) - 2\alpha (e\Phi -2)]\delta^{ij}
\end{array} 
\right),
\end{eqnarray} 
where
\begin{eqnarray}
\mathcal{Q}=(1+c^2_s-v^2) +2\alpha (e\vec{v}\cdot\vec{\Lambda}-3v^2-e\Phi +3) +4\alpha c^2_s.
\end{eqnarray}
Note that we have obtained a modified relativistic acoustic metric with new terms emerging. Especially the terms generated due to the scalar potential $\Phi$ and the vector potential $\vec{\Lambda}$.
{Furthermore, in the absence of potentials, i.e., by taking $e=0$, we find the relativistic acoustic metric with GUP corrections, given by
\begin{eqnarray}
g_{\mu\nu}=\frac{b\rho_0}{c_s\sqrt{\mathcal{Q}}} 
\left(\begin{array}{ccc}
-c^2_s(1+2\alpha) + v^2(1+4\alpha) & \vdots & -v^{i}(1+4\alpha) \\ 
\cdots\cdots &\cdot &\cdots\cdots\\
-v^{j}(1+4\alpha) & \vdots & [1+4\alpha + c^2_s(1+2\alpha)]\delta^{ij}
\end{array} 
\right),
\end{eqnarray}
where
\begin{eqnarray}
\mathcal{Q}=(1+c^2_s-v^2) +2\alpha (3 +2c^2_s -3v^2).
\end{eqnarray}
In addition, for $ \alpha=0 $, we have
\begin{eqnarray}
g_{\mu\nu}=\frac{b\rho_0}{c_s\sqrt{1+c^2_s-v^2}} 
\left(\begin{array}{ccc}
-c^2_s + v^2 & \vdots & -v^{i} \\ 
\cdots\cdots &\cdot &\cdots\cdots\\
-v^{j} & \vdots & (1+c^2_s)\delta^{ij}
\end{array} 
\right),
\end{eqnarray}
such that the relativistic acoustic metric obtained in~\cite{Ge:2010wx} is recovered.}
However, in the non-relativistic limit the metric (\ref{metr}) becomes
\begin{eqnarray}
\label{metnr}
g_{\mu\nu}=\frac{b\rho_0}{c_s\sqrt{f}} 
\left(\begin{array}{ccc}
-c^2_s(1+2\alpha) + v^2(1+4\alpha)- 2\alpha e\vec{v}\cdot\vec{\Lambda} & \vdots & -v^{i}(1+4\alpha 
-\alpha e\Phi)+\alpha e\Lambda^i \\ 
\cdots\cdots &\cdot &\cdots\cdots\\
-v^{j}(1+4\alpha -\alpha e\Phi)+\alpha e\Lambda^j & \vdots & [1- 2\alpha (e\Phi -2)]\delta^{ij}
\end{array} 
\right),
\end{eqnarray} 
where
\begin{eqnarray}
f=1+6\alpha +2\alpha (e\vec{v}\cdot\vec{\Lambda}-e\Phi).
\end{eqnarray}
{So for $ e=0 $ in the metric above, we find}
\begin{eqnarray}
g_{\mu\nu}=\frac{b\rho_0}{c_s\sqrt{1+6\alpha}} 
\left(\begin{array}{ccc}
-c^2_s(1+2\alpha) + v^2(1+4\alpha) & \vdots & -v^{i}(1+4\alpha) \\ 
\cdots\cdots &\cdot &\cdots\cdots\\
-v^{j}(1+4\alpha) & \vdots & \left(1+4\alpha\right)\delta^{ij}
\end{array} 
\right),
\end{eqnarray}
{which is the metric of the acoustic black hole with GUP corrections without the effect of potentials.

Now, for $\alpha=0$, we obtain the metric found by Unruh~\cite{Unruh:1980cg} up to an overall factor}
\begin{eqnarray}
g_{\mu\nu}=\frac{b\rho_0}{c_s} 
\left(\begin{array}{ccc}
-(c^2_s - v^2) & \vdots & -v^{i} \\ 
\cdots\cdots &\cdot &\cdots\cdots\\
-v^{j} & \vdots & \delta^{ij}
\end{array} 
\right).
\end{eqnarray}
We can also write the effective acoustic line element for the metric (\ref{metnr}) as follows:
\begin{eqnarray}
ds^2&=&\frac{b\rho_0}{c_s\sqrt{f}}
\left[-F(v)\,dt^2 
- 2\vec{V}\cdot d\vec{x}dt
+[1- 2\alpha (e\Phi -2)] dx^2\right],
\end{eqnarray}
where 
\begin{eqnarray}
&&F(v)=c^2_s(1+2\alpha) - v^2(1+4\alpha)+ 2\alpha e\vec{v}\cdot\vec{\Lambda},
\\
&&\vec{V}=(1+4\alpha -\alpha e\Phi)\vec{v}-e\alpha\vec{\Lambda}.
\end{eqnarray}
Now, by performing a coordinate change
\begin{eqnarray}
d\tau=dt + \frac{\vec{V}\cdot d\vec{x}}{F(v)},
\end{eqnarray}
the line element can be written in stationary form
\begin{eqnarray}
\label{elsf}
ds^2=\frac{b\rho_0}{c_s\sqrt{f}}
\left[-F(v)\,d\tau^2 
+\left(\frac{V^i V^j}{F(v)}+\left[1- 2\alpha (e\Phi -2)\right]\delta^{ij}\right) dx^idx^j\right].
\end{eqnarray}

At this point, to address the Hawking temperature issue, we will consider the configuration of an incompressible fluid with spherical symmetry. 
For this situation  the density $ \rho $ is a position independent quantity, the sound speed is also a constant and the continuity equation implies that $ v\sim 1/r^2 $.
Thus, disregarding position-independent factors, the line element (\ref{elsf}) written in the Schwarzschild-type form is given by
\begin{eqnarray}
\label{canmet}
ds^2=-\mathcal{F}(v_r)d\tau^2 + \frac{c^2_s\left(1+6\alpha -2\alpha e\Phi\right)}
{f(v_r)\mathcal{F}(v_r)}dr^2 
+ \frac{\left[1- 2\alpha (e\Phi -2)\right]}{\sqrt{f(v_r)}}r^2d\Omega^2,
\end{eqnarray}
where
\begin{eqnarray}
\mathcal{F}(v_r)=\frac{F(v_r)}{\sqrt{f(v_r)}}
\approx c^2_s[1-\alpha(1-e\Phi + e v_r\Lambda_r)]
-v^2_r[1-\alpha(ev_r\Lambda_r -e\Phi -1 ) ] + 2e\alpha v_r\Lambda_r.
\end{eqnarray}
Next, by continuity equation $ v_r=c_s r^2_h/r^2 $, where $ r_h $ is the radius of the event horizon, 
{we have that the modified canonical acoustic black hole reads}
\begin{eqnarray}
\label{fmt}
\mathcal{F}(r)=
c^2_s\left[1-\alpha\left(1-e\Phi(r) + \frac{e c_s r^2_h \Lambda_r(r)}{r^2}\right)\right]
-\frac{c^2_s r^4_h}{r^4}\left[1-\alpha\left(\frac{e c_s r^2_h \Lambda_r(r)}{r^2} -e\Phi(r) -1\right)\right] 
+ \frac{2e\alpha c_s r^2_h \Lambda_r(r)}{r^2}.
\end{eqnarray}
{Notice that when $ e=0 $ and $ c_s=1 $ the metric function above is given by
\begin{eqnarray}
\mathcal{F}(r)=1 - \frac{r^4_h}{r^4},
\end{eqnarray}
which is the canonical acoustic black hole metric found by Visser~\cite{Visser:1997ux}. 
This metric is different from the well-known four-dimensional black hole geometries normally found in general relativity. 
To obtain a result conforming to the Schwarzschild geometry, but not identical to it,  we can consider from the continuity equation that $\rho v_r\propto 1/r^2$ with $ v_r=\sqrt{2GM/r} $ and $\rho \propto r^{-3/2}$, such a way for the metric function, we find that $ \mathcal{F}(r)=1- 2GM/r $ --- see~\cite{Visser:1997ux} for further details.}

Therefore, the Hawking temperature in this case is given by
\begin{eqnarray}
\tilde{T}_H&=&\frac{\mathcal{F}^{\prime}(r_h)}{4\pi}
\\
&=&\frac{c^2_s(1+\alpha)+e\alpha c^2_s\left[\Phi(r_h)- c_s\Lambda_r(r_h)\right]}{\pi r_h}
+\frac{2e\alpha c_s}{4\pi}\left(\Lambda^{\prime}_r(r_h)-\frac{2\Lambda_r(r_h)}{r_h}\right).
\label{ht}
\end{eqnarray}
Note that the Hawking temperature of the modified canonical acoustic black hole is affected by the GUP deformation parameter $\alpha$ and also by the contribution of the potentials $\Phi$ and $\Lambda_r$. 
In particular, for $ \Lambda_r=0 $ the Hawking temperature in this case is reduced to
\begin{eqnarray}
\tilde{T}_H=\frac{c^2_s[1+\alpha + e\alpha\Phi(r_h)]}{\pi r_h}.
\end{eqnarray}
Furthermore, assuming $ c_s=1 $ and a scalar potential of the type $\Phi=r_0/r$, the temperature becomes
\begin{eqnarray}
\label{temp}
\tilde{T}_H=T_H\left(1+\alpha + \frac{\alpha e r_0}{r_h}\right),
\end{eqnarray}
where $ T_H=1/\pi r_h $ is the Hawking temperature of the canonical  acoustic black hole for $ \alpha=0 $ and $r_0$ is a parameter with length dimension.
{Therefore, in the absence of the scalar potential  ($r_0=0$ or $e=0$), we have
\begin{eqnarray}
\label{tgup}
\tilde{T}_H=T_H\left(1+\alpha\right),
\end{eqnarray}
and so the Hawking temperature has its value increased when we vary the parameter $\alpha$. On the other hand, when the scalar potential is considered the temperature is modified and a new correction term appears. }

Furthermore, with the result obtained for the temperature in (\ref{temp}) we can determine 
the entropy (entanglement entropy~\cite{Anacleto:2019rfn}) of the modified canonical acoustic black hole 
by applying the first law of thermodynamics such that
\begin{eqnarray}
S&=&\int \frac{dE}{dT}=\int \frac{\kappa\, dA}{8\pi\tilde{T}_H}
=\int \frac{dA}{4}\left(1-\alpha^2 - \frac{2\sqrt{\pi}\alpha^2 e r_0}{\sqrt{A}} - \frac{4\pi\alpha^2 e^2 r^2_0}{A}\right),
\\
&=&\frac{(1-\alpha^2)A}{4} - \frac{4\sqrt{\pi}\alpha^2 e r_0\sqrt{A}}{4} - \frac{4\pi\alpha^2 e^2 r^2_0}{4}\ln\left(\frac{A}{4r^2_0}\right),
\end{eqnarray}
where $ A=4\pi r^2_h $ is the horizon area of the canonical  acoustic black hole.
Note that we have found a logarithmic correction term for entropy. This correction term is generated due to the GUP and the scalar potential. 
The presence of the logarithmic correction term is related to the appearance of black hole remnants in gravitation~\cite{Nozari:2005ah}.

The metric~\eqref{canmet} can also be written, for $\Lambda_r=0$, as follows
\begin{eqnarray}
ds^2=-\mathcal{F}(v_r)d\tau^2 + \frac{c^2_s}
{\mathcal{F}(v_r)}dr^2 
+ \left(1+\alpha-\alpha e\Phi\right)r^2d\Omega^2,
\end{eqnarray}
where
\begin{eqnarray}
\mathcal{F}(v_r)= c^2_s[1-\alpha(1-e\Phi)]
-v^2_r[1+\alpha(1+e\Phi) ].
\end{eqnarray}
Now, recalling that $ v_r=c_s r^2_h/r^2 $ and $\Phi=r_0/r$ and making $c_s=1 $, we find
\begin{eqnarray}
ds^2=-\mathcal{F}(r)d\tau^2 + \frac{dr^2}{\mathcal{F}(r)} 
+ \left(1+\alpha-\frac{\alpha er_0}{r}\right)r^2d\Omega^2,
\end{eqnarray}
being
\begin{eqnarray}
\mathcal{F}(r)=1-\alpha + \frac{\alpha er_0}{r}
-\frac{r^4_h}{r^4}\left[1+\alpha\left(1+\frac{er_0}{r}\right) \right].
\end{eqnarray}
At the limit of $ r\gg r_h\equiv \alpha er_0 $ the metric becomes
\begin{eqnarray}
ds^2=-\mathcal{F}(r)d\tau^2 + \frac{dr^2}{\mathcal{F}(r)} + r^2d\Omega^2,
\end{eqnarray}
that in this case the metric function $\mathcal{F}(r)$ is now given by
\begin{eqnarray}
\mathcal{F}(r)=1-\alpha^2 + \frac{r_h}{r},
\end{eqnarray}
which is an asymptotically global monopole metric.
So this result is a consequence of the scalar potential effect on the effective acoustic metric.

{Now, for the case $ \Phi=0 $ and $ \Lambda=\lambda/r $, being $ \lambda $ a parameter with a length dimension. 
In this case the Hawking temperature in~\eqref{ht} becomes
\begin{eqnarray}
\label{temp2}
\tilde{T}_H=T_H\left(1+\alpha - \frac{5\alpha e\lambda}{2r_h}\right),
\end{eqnarray}
and for entropy we find
\begin{eqnarray}
S&=&\frac{(1-\alpha^2)A}{4} + \frac{20\sqrt{\pi}\alpha^2 e \lambda\sqrt{A}}{4} - \frac{25\pi\alpha^2 e^2 \lambda^2}{4}\ln\left(\frac{A}{4\lambda^2}\right).
\end{eqnarray}
Again a logarithmic correction is also generated for the pure vector potential contribution whose pre-factor with respect to the pure scalar potential case is $(5\lambda/2r_0)^2$, which mimics logarithm corrections due to species with different spins.

Here it is worth mentioning that considering the contributions of $\Phi=r_0/r$ and $\Lambda_r=\lambda/r$ 
the metric function~\eqref{fmt} becomes
\begin{eqnarray}
\label{fmt2}
\mathcal{F}(r)=
1-\alpha\left(1-\frac{e r_0}{r} - \frac{e\lambda r^2_h}{r^3}\right)
-\frac{r^4_h}{r^4}\left[1+\alpha\left(1+\frac{e r_0}{r}-\frac{e\lambda r^2_h}{r^3}\right)\right].
\end{eqnarray}
{From (\ref{ht}), by making $ c_s=1 $, $\Phi=r_0/r$ and $\Lambda_r=\lambda/r$, we obtain the Hawking temperature which is given by}
\begin{eqnarray}
\tilde{T}_H
=T_H\left(1+\alpha + \frac{e\alpha r_0}{r_h}-\frac{5e\alpha\lambda}{2r_h}\right).
\end{eqnarray}
On the other hand, if $\Lambda_r=\lambda_0$, where $\lambda_0$ is a dimensionless constant, then we have
\begin{eqnarray}
\label{fmrn}
\mathcal{F}(r)=
1-\alpha\left(1-\frac{e r_0}{r} - \frac{e\lambda_0 r^2_h}{r^2}\right)
-\frac{r^4_h}{r^4}\left[1+\alpha\left(1+\frac{e r_0}{r}-\frac{e\lambda_0 r^2_h}{r^2}\right)\right].
\end{eqnarray}
{Our result shows similarity with the metric obtained in~\cite{Ling:2021vgk} 
with the identification of effective charge and  mass as $ Q^2=\alpha e\lambda_0 r^2_h $ and $ 2M=\alpha e r_0 $ for $ r_0<0 $, respectively.}
Therefore, we highlight another new result that arises due to the GUP effect, that is, the contributions of $\Phi$ and $\Lambda$ can generate Reissner-Nordstr\"om-like contributions for the metric.}
{In this case the Hawking temperature is given by equation (\ref{ht}) making $ c_s=1 $, $\Phi=r_0/r$ and $\Lambda_r=\lambda_0$. So we have
\begin{eqnarray}
\tilde{T}_H
=T_H\left(1+\alpha -2e\alpha\lambda_0 + \frac{e\alpha r_0}{r_h}\right).
\end{eqnarray}
Hence, when $e=0$, that is, $Q=M=0$, we have the temperature given by equation (\ref{tgup}).}

\section{The Dispersion Relation}
\label{dispersion}

In order to study further effects from the GUP we shall now investigate the dispersion relation.
For this we will use the following notation:
\begin{eqnarray}
S_1\sim \mbox{Re}\left[e^{i\tilde{\omega} t - i \vec{k}\cdot\vec{x}}\right],
\qquad
\tilde{\omega}=\frac{\partial S_1}{\partial t},
\qquad
\vec{k}=\nabla S_1.
\end{eqnarray}
In this way the Klein-Gordon equation \eqref{KG} in terms of momenta and frequency becomes
\begin{eqnarray}
a\,\tilde{\omega}^2 + b\,\tilde{\omega} + c = 0,
\end{eqnarray}
where
\begin{eqnarray}
&&a=1- 2\alpha(e\Phi-2)+ c_s^2\left( 1+2\alpha\right),
\\
&&b=2(1+4\alpha-\alpha e\Phi)(\vec{v}\cdot\vec{k})- 2\alpha (\vec{\Lambda}\cdot\vec{k}),
\\
&&c=-\left[-(1+4\alpha) v^2 k^2 + 2\alpha e(\vec{\Lambda}\cdot\vec{k})( \vec{v}\cdot\vec{k} )
+c_s^2\left( 1+2\alpha\right)k^2 \right].
\end{eqnarray}
Now, by considering $ k^i=\delta^{i1} $, we have
\begin{eqnarray}
\tilde{\omega}=\frac{-2\left[(1+4\alpha + \alpha e\Phi)v_1 -\alpha e\Lambda_1  \right]k\pm \sqrt{\Delta}}{2a},
\end{eqnarray}
with 
\begin{eqnarray}
\sqrt{\Delta}
=2c_sk\sqrt{1+2\alpha -2\alpha(e\Phi - 2) + 4\alpha e\Phi\frac{v_1^2}{c^2_s}
-{2\alpha\frac{e\Lambda_1 v_1 }{c^2_s} }
+\left[(1+4\alpha)c^2_s - (1+6\alpha)v^2_1 + {2\alpha e\Lambda_1 v_1 }\right] }.
\end{eqnarray}
In particular for $ \Lambda_1=0 $ and in the non-relativistic limit, we obtain
\begin{eqnarray}
\tilde{\omega} \approx \pm \frac{c_s\sqrt{1+6\alpha - 2\alpha e\Phi}}
{1+4\alpha -2\alpha e\Phi}k
=\eta\, \omega\left(1-\alpha + \alpha e\Phi\right),
\end{eqnarray}
being $ \omega=c_s k$ (the dispersion relation for $ \alpha=0 $) and 
$ \eta=\pm 1 $ the polarizations.
Now, to explore the scalar potential effect recall that
$ \Phi=r_0/r $. Hence, focusing on the dispersion relation in the vicinity of the event horizon, 
$ r \rightarrow r_h $, becomes
\begin{eqnarray}
\tilde{\omega}=\eta\,\omega\left(1-\alpha + \frac{er_0\alpha}{r_h}\right)
=\eta(1-\alpha)\omega \left(1+ \frac{er_0\alpha}{r_h}\right) + \mathcal{O}(\alpha^2 ).
\end{eqnarray}
Next, we redefine $ \bar{\omega}=\tilde{\omega}/(1-\alpha) $ 
and apply $ k\sim \Delta k \geq 1/\Delta x=1/r_h $.
Thus, the dispersion relation above is rewritten in the form
\begin{eqnarray}
\label{drm}
\bar{\omega}=\eta\,\omega \left(1+ \alpha er_0 k\right).
\end{eqnarray}
So we note that the scalar potential contribution generated by the GUP is associated with a viscosity effect in the fluid. Similar behavior has been analyzed in~\cite{Anacleto:2019rfn} in the background with Lorentz violation. 

We can also write an expression in terms of the frequency difference given by
\begin{eqnarray}
\frac{\Delta\omega}{\omega}=\frac{\bar{\omega}-\eta\omega}{\omega}=\eta \alpha er_0 k.
\end{eqnarray}
Now we can get from the dispersion relation (\ref{drm}) the phase velocity ($v_p$) and the group velocity ($v_g$) as follows:
\begin{eqnarray}
v_p=\frac{\bar{\omega}}{k}=\eta c_s(1+\alpha er_0 k),
\end{eqnarray}
and 
\begin{eqnarray}
v_g=\frac{d\bar{\omega}}{dk}=\eta c_s(1+2\alpha er_0 k).
\end{eqnarray}
Using the Rayleigh's formula
\begin{eqnarray}
v_g=v_p + k\frac{dv_p}{dk},
\end{eqnarray}
we obtain an expression for the velocity difference given by
\begin{eqnarray}
\frac{v_g - v_p}{v_p}=\alpha er_0 k.
\end{eqnarray}
In this case, 
we have a supersonic behavior ($ v_g >v_p $) as in BEC systems whose excitations follow `superluminal' dispersion relations \cite{Robertson:2012ku}.

Furthermore, from equation (\ref{temp}), for the Hawking temperature we can also obtain a relationship for the temperature difference. So we have
\begin{eqnarray}
\mathcal{T}_H=T_H(1+ \alpha er_0 k).
\end{eqnarray}
Here we have redefined $ \mathcal{T}_H=\tilde{T}_H/(1+\alpha) $.
Hence, we find the following relationship for the Hawking temperature variation due to the potential effect
\begin{eqnarray}
\frac{\Delta T_H}{T_H}=\frac{\mathcal{T}_H-T_H}{T_H}=\alpha er_0 k.
\end{eqnarray}
Therefore, we find the following relationship:
\begin{eqnarray}
\frac{v_g - v_p}{v_p}=\eta\frac{\Delta\omega}{\omega}=\frac{\Delta T_H}{T_H}.
\end{eqnarray}
Moreover, we emphasize that the modification of the dispersion relation associated with $\delta=\alpha er_0 k$ is analogous to Lorentz-violating parameter in BEC physics whose deviation is $\delta\sim 10^{-5}$ --- see ~\cite{Anacleto:2019rfn} and references therein. As a result, by recalling $k\sim 1/r_h$, such that $\delta\sim\alpha e\Phi(r_h)$, we obtain
\begin{eqnarray}
\frac{\Delta T_H}{T_H}\sim \Phi(r_h)\sim 10^{-5}.
\end{eqnarray}
Interestingly enough, this result is also analogous to the small temperature fluctuations in the anisotropy of the cosmic microwave
background (CMB), with the scalar field evaluated at horizon corresponding to fluctuations of the gravitational potential in the cosmological scenario.

\section{Conclusions}
\label{conclu}

We obtained an effective acoustic metric for acoustic black holes with quantum corrections 
implemented by the generalized Heisenberg uncertainty principle (GUP) in the Abelian Higgs model. Thus, contrary to the usual acoustic metrics, the obtained effective acoustic metric now depends on the contribution of scalar and vector potentials that emerge as an effect of the GUP.
We have examined the Hawking temperature and the entropy for the modified canonical acoustic black hole. As a result, the Hawking temperature that changes due to the effect of the GUP provides logarithmic corrections to the entropy. We address the modification of the Hawking temperature by considering different effects by turning on scalar and vector potential contributions separately and also turning on them together.
Moreover, by analyzing the dispersion relation of the model we have obtained a relationship between the group and phase velocity difference due to the GUP effect, as well as expressions for the frequency and temperature deviations. Finally, we have made a few comments on how they can be related to BEC physics with supersonic dispersion relations and small temperature fluctuations in the anisotropy of the cosmic microwave background (CMB).

\acknowledgments

We would like to thank CNPq, CAPES and CNPq/PRONEX/FAPESQ-PB (Grant nos. 165/2018 and 015/2019),
for partial financial support. MAA, FAB and EP acknowledge support from CNPq (Grant nos. 306962/2018-7 and
433980/2018-4, 312104/2018-9, 304852/2017-1).


\end{document}